\documentclass[nopreprintline]{elsart}

\pdfoutput=1

\newcounter{bla}

\newcommand{\bea}{\begin{eqnarray}}
\newcommand{\eea}{\end{eqnarray}\noindent}
\newcommand{\be}{\begin{equation}}
\newcommand{\ee}{\end{equation}}
\newcommand{\nn}{\nonumber}
\newcommand{\calst}{\mbox{$\cal S$}}

\newcommand{\diff}[1][{}]{{\mathrm{d}}^{#1}\!}
\newcommand{\bul}{\mbox{\bf\textperiodcentered}}
\newcommand{\tr}[2][{}]{tr^{#1}\!\left\{{#2}\right\}}
\newcommand{\GOLEMC}{{\texttt{go\-lem95C}}}
\def\eps{\epsilon}
\usepackage{amssymb,amsmath}
\usepackage{epsfig}
\usepackage{listings}

\begin{document}

\bibliographystyle{unsrt}

\begin{frontmatter}

\hfill{LAPTH-072/13}\\
\hfill{MPP-2013-314}

%\title{Tools for spin two particles: extension of the golem95C integral library}
\title{Tools for NLO automation: extension of the golem95C integral library}

\author[a]{J.~Ph.~Guillet},
\author[b]{G.~Heinrich},
\author[b]{J.~F.~von Soden-Fraunhofen}

\address[a]{LAPTH, Universit\'e de Savoie and CNRS, Annecy-le-Vieux, France}
\address[b]{Max-Planck-Institut f\"ur Physik, F\"ohringer Ring 6, 80805 M\"unchen, Germany}
%\address[c]{Institute for Particle Physics Phenomenology,
%        University of Durham, \\Durham, DH1 3LE, UK}

%\thanks{corresponding author: G.~Heinrich}

\begin{abstract}
We present an extension of the program golem95C 
for the numerical evaluation of scalar 
integrals and tensor form factors 
entering the calculation of one-loop amp\-litudes, 
which supports tensor ranks exceeding the number of propagators. 
This extension allows various applications in Beyond the Standard Model physics 
and effective theories,
for example higher ranks due to propagators of spin two particles, 
or due to effective vertices. Complex masses are also supported.
The program is not restricted to the Feynman diagrammatic approach, 
as it also contains routines to interface to unitarity-inspired 
numerical reconstruction of the integrand at the tensorial level.
Therefore it can serve as a general integral library in automated programs to 
calculate one-loop amplitudes.
\begin{flushleft}
\end{flushleft}

\begin{keyword}
NLO computations, one-loop diagrams, tensor integrals, higher spin representations, effective theories
  % Please give some freely chosen keywords that we can use in a
  % cumulative keyword index.
\end{keyword}

\end{abstract}

\end{frontmatter}

\newpage

{\bf NEW VERSION PROGRAM SUMMARY}

\begin{small}
\noindent
{\em Manuscript Title: Tools for NLO automation: extension of the golem95C integral library}                                       \\
{\em Authors: J.~Ph.~Guillet, G.~Heinrich, J.~F.~von Soden-Fraunhofen}                                                \\
{\em Program Title: golem95-1.3.0}                                          \\
%{\em Journal Reference:}                                      \\
  %Leave blank, supplied by Elsevier.
%{\em Catalogue identifier:}                                   \\
  %Leave blank, supplied by Elsevier.
{\em Licensing provisions:}      none                             \\
  %enter "none" if CPC non-profit use license is sufficient.
{\em Programming language:}     Fortran95                              \\
{\em Computer: Any computer with a  Fortran95 compiler }   \\
  %Computer(s) for which program has been designed.
{\em Operating system:}      Linux, Unix                                   \\
  %Operating system(s) for which program has been designed.
{\em RAM:} RAM used per integral/form factor  is insignificant                                             \\
  %RAM in bytes required to execute program with typical data.
%{\em Number of processors used:}      one                        \\
  %If more than one processor.
%{\em Supplementary material:}                                 \\
  % Fill in if necessary, otherwise leave out.
  % Please give some freely chosen keywords that we can use in a
  % cumulative keyword index.
  % see http://www.aip.org/pacs/pacs.html
{\em Classification:} 4.4, 11.1                                       \\
  %Classify using CPC Program Library Subject Index, see (
  % http://cpc.cs.qub.ac.uk/subjectIndex/SUBJECT_index.html)
  %e.g. 4.4 Feynman diagrams, 5 Computer Algebra.
{\em External routines/libraries:}   some finite scalar integrals are called from OneLOop\,\cite{vanHameren:2009dr,vanHameren:2010cp}, 
 the option to call them from LoopTools\,\cite{Hahn:1998yk,Hahn:2010zi} is also implemented.                               \\
  % Fill in if necessary, otherwise leave out.
%{\em Subprograms used:}                                       \\
  %Fill in if necessary, otherwise leave out.
{\em Catalogue identifier of previous version:}   AEEO\_v2\_0           \\
  %Only required for a New Version summary, otherwise leave out.
{\em Journal reference of previous version:}    Comput. Phys. Commun. 182 (2011) 2276.   \\
  %Only required for a New Version summary, otherwise leave out.
{\em Does the new version supersede the previous version?:} yes   \\
  %Only required for a New Version summary, otherwise leave out.
{\em Nature of problem:} Evaluation of 
one-loop multi-leg integrals occurring in the calculation of next-to-leading order corrections
to scattering amplitudes in particle physics. In the presence of 
particles with spin two in the loop, or effective vertices, or certain gauges, 
tensor integrals where the rank exceeds the number of propagators $N$ are required.
\\
{\em Solution method:} Extension of the reduction algorithm to rank $r\leq 10$ 
for $N\leq 4$ and $r\leq N+1$ for $N\geq 5$,
 which is sufficient for most applications in Beyond the Standard Model Physics. \\
   {\em Reasons for the new version:} The previous version was restricted to 
   tensor ranks less or equal to the number of propagators.\\
   {\em Summary of revisions:} Tensor ranks $r>N$ are supported,
   an alternative reduction method for the case of small Gram determinants is implemented,
   numerical stability for the case of small mass differences has been improved.\\
{\em Running time:} Depends on the nature of the problem. 
A single call to a rank 6 five-point form factor 
at a randomly chosen kinematic point, using real masses,  takes $10^{-3}$ seconds on an 
Intel Core 4 i7-3770 CPU with a 3.4\,GHz processor. \\
%{\em References:}
%\begin{refnummer}
%\item A. van Hameren, JHEP 0909:106,2009.         
%\item T. Hahn, M. Perez-Victoria, Comput.\ Phys.\ Commun.\  {\bf 118} (1999) 153.
% This is the reference list of the Program Summary
% Type references in text as [1], [2], etc.
% This list is different from the bibliography, which you can use in the Long Write-Up.
%\end{refnummer}

\end{small}

\newpage

%\hspace{1pc}

\section{Introduction}

Over the last  years, enormous progress has been made to 
push the calculation of NLO corrections towards a higher number of 
particles in the final states, i.e. to ``multi-leg" amplitudes, 
in QCD as well as in the electroweak sector. 

Nowadays, the efforts  are  focused on the goals of {\em automating} 
multi-leg one-loop calculations and making them {\em publicly available}. 
Programs aiming at the complete automation of one-loop amplitude calculations, 
including the generation of the amplitude requested by the user ``on the fly", 
are e.g. {\sc Feyn\-Arts/FormCalc}\,\cite{Hahn:1998yk,Agrawal:2012cv}, 
{\tt HELAC-NLO}\,\cite{Bevilacqua:2011xh}, {\sc GoSam}\,\cite{Cullen:2011ac}, 
{\tt aMC@NLO}\,\cite{Hirschi:2011pa},
{\tt BlackHat}\,\cite{Berger:2008sj}, {\tt NJET}\,\cite{Badger:2012pf},
{\tt OpenLoops}\,\cite{Cascioli:2011va}, 
{\tt Recola}\,\cite{Actis:2012qn}.
%where the latter two are not yet publicly available.
Public programs which contain a collection of pre-generated processes 
%rather than producing the amplitude ``on the fly",
are e.g. {\tt MCFM}\,\cite{Campbell:2002tg}, {\tt VBFNLO}\,\cite{Arnold:2008rz}.

An important ingredient for such programs is an integral library containing the one-loop 
integrals which are the basic building blocks of  any one-loop amplitude unless it is
calculated purely numerically. 
Several libraries are publicly available to date: 
{\tt FF}\,\cite{vanOldenborgh:1990yc}, {\tt Looptools}\,\cite{Hahn:2010zi}, 
{\tt QCDLoop}\,\cite{Ellis:2007qk}, 
{\tt OneLOop}\,\cite{vanHameren:2010cp},  
{\tt golem95C}\,\cite{Binoth:2008uq,Cullen:2011kv}, {\tt PJFRY}\,\cite{Fleischer:2011zz}.
%Hexagon.F\,\cite{Diakonidis:2010rs}.
A code for the calculation of one-loop four-point functions with complex masses ({\tt D0C}) 
can be found in \cite{Nhung:2009pm}.  The latter has been integrated into the {\tt LoopTools} 
library\,\cite{Hahn:2010zi} where the complex versions of  infrared finite 
integrals with less than four legs are already implemented. 
A complete set of scalar four-point integrals, both in dimensional and in mass regularisation 
and valid also for complex masses 
can be found in \cite{Denner:2010tr} in analytic form. 

Public programs dedicated to the reduction 
of  multi-leg one-loop amplitudes at integrand level are e.g. 
{\tt CutTools}\,\cite{Ossola:2007ax} and
{\tt Samurai}\,\cite{Mastrolia:2010nb}. A novel reduction algorithm 
based on integrand reduction through Laurent series expansion
also has been developed~\cite{Mastrolia:2012bu}.

The calculation of scalar one-loop integrals has a long tradition of pioneering work, see e.g. 
\cite{vanOldenborgh:1989wn,tHooft:1978xw,Fabricius:1979tb,Beenakker:1988jr,Denner:1991qq,Bern:1992em,Bern:1993kr}.  
For processes involving unstable particles, these integrals are also required for 
complex internal masses, in order to be able to work within the so-called 
``complex-mass scheme" developed in Refs.\,\cite{Denner:1999gp,Denner:2005fg}.
An extension of the {\tt golem95} library to complex masses has been presented 
in Ref.\,\cite{Cullen:2011kv}, called \GOLEMC{}.
The strategy in \GOLEMC{} to avoid numerical instabilities due to small Gram determinants 
is to avoid the reduction to scalar basis integrals in such cases, in favour of turning to 
the numerical evaluation of a convenient one-dimensional integral representation of the 
respective tensor integral.
Recent new developments in this direction can be found in Ref.~\cite{Guillet:2013mta}.

In this article, we present an extension  of the {\tt golem95C} library 
to integrals with tensor ranks $r$ exceeding the number of propagators $N$.
Such integrals occur for example in effective theories (a prominent example is the 
effective coupling of gluons to the Higgs boson), or in calculations within
theories containing spin two particles beyond the leading order.  
%a procedure how to treat higher ranks is also mandatory.
Even though most of the modern methods to reduce one-loop amplitudes do not entirely 
rely on Feynman diagrams and tensor reduction anymore, the latter is still important in a number of cases. 
In particular, methods based on the reconstruction of  tensor coefficients at 
integrand level \,\cite{vanHameren:2009vq,Heinrich:2010ax}
have  proven very efficient recently~\cite{Cascioli:2011va,Actis:2012qn}.
%For example, for final states of lower multiplicity, tensor reduction, supplemented with a 
%well established treatment of small Gram determinants, 
%is still competitive in speed and  robustness  with unitarity inspired methods. 
%universal applicability 
Therefore, a tensor and scalar integral library represents an
important tool for calculations within and beyond the Standard Model. 

Further, tensor reduction, or tensorial reconstruction\,\cite{Heinrich:2010ax}, can be an important 
``rescue system" for phase space points where cut-based techniques do not provide the 
desired accuracy. As such, the {\tt golem95C} library is an important ingredient 
for the automated one-loop program {\sc GoSam}\,\cite{Cullen:2011ac}.
The extension of cut based reduction methods at integrand level to higher ranks 
also has been tackled already\,\cite{Mastrolia:2012bu}. 
The  higher rank integrals 
(closely related to integrals in more than $D=4-2\eps$ space-time dimensions) 
can be written in terms of some ``basis integrals" which 
can be called from the {\tt golem95C} library presented here.
However, the library contains the higher rank form factors in full generality. 
Therefore it can be used in various applications where the number of loop momenta
in the integrand is large, 
in combination with integrand reduction methods ranging from traditional tensor reduction to 
cut-based methods.

This article is organized as follows. In Section 2, we review the theoretical background, 
while in Section \ref{sec:highrank} we focus on the new extension of the library to integrals 
of higher ranks. Section 4 contains installation instructions, while in Section 
\ref{sec:examples} the user can 
find simple examples how to run the program. Section 6 contains our conclusions.

\section{Theoretical background}

The program is an update of the tensor and scalar integral 
library described in more detail in  Ref.~\cite{Binoth:2008uq}, 
based on the formalism  developed in Refs.~\cite{Binoth:2005ff,Binoth:1999sp}
to reduce tensor integrals to a convenient set of basis integrals. 
Similar reduction schemes can be found e.g. in 
Refs.~\cite{Bern:1992em,Bern:1993kr,Duplancic:2003tv,Giele:2004iy,delAguila:2004nf,vanHameren:2005ed,Denner:2005nn,Fleischer:2010sq,Fleischer:2011hc}.
Here we will describe the theoretical framework only  briefly 
and focus on the new features of the program.

\subsection{Form Factors}

Tensor integrals can be divided into a part containing the Lorentz structure 
and a part consisting of scalar quantities, which we call {\it form factors}.

We define an $N$-point tensor integral of rank $r$ in $D=4-2\eps$ 
dimensions as 
\begin{eqnarray}
I^{D,\,\mu_1\ldots\mu_r}_N(a_1,\ldots,a_r) = 
\int \frac{d^D q}{i \, \pi^{D/2}}
\; \frac{q_{a_1}^{\mu_1}\,\dots  q_{a_r}^{\mu_r}}{
(q_1^2-m_1^2+i\delta)\dots (q_N^2-m_N^2+i\delta)}
\label{eq0}
\end{eqnarray} 
where $q_a=q+r_a$, $q$ is the loop momentum, and 
$r_a$ is a combination of external momenta.
Using the shift invariant vectors 
\begin{equation}
\Delta_{ij}^\mu=
r_i^\mu - r_j^\mu\;,
\end{equation}
we can write down a general form factor decomposition of an arbitrary tensor integral
\begin{eqnarray}\label{eq:formfactordef}
I^{D,\mu_1\ldots\mu_r}_N(a_1,\ldots, a_r; S)&=&
\sum_{j_1,\ldots,j_r\in S}
   \left[\Delta_{j_1\cdot}^{\cdot}\cdots\Delta_{j_r\cdot}^{\cdot}%
   \right]^{\{\mu_1\ldots\mu_r\}}_{\{a_1\ldots a_r\}}
   A^{N,r}_{j_1\ldots j_r}(S)\\
&&+ \sum_{j_1,\ldots,j_{r-2}\in S}
   \left[g^{\cdot\cdot}%
   \Delta_{j_1\cdot}^{\cdot}\cdots\Delta_{j_{r-2}\cdot}^{\cdot}%
   \right]^{\{\mu_1\ldots\mu_r\}}_{\{a_1\ldots a_r\}}
   B^{N,r}_{j_1\ldots j_{r-2}}(S)
\nn\\
&&+ \sum_{j_1,\ldots,j_{r-4}\in S}
   \left[g^{\cdot\cdot}g^{\cdot\cdot}%
   \Delta_{j_1\cdot}^{\cdot}\cdots\Delta_{j_{r-4}\cdot}^{\cdot}%
   \right]^{\{\mu_1\ldots\mu_r\}}_{\{a_1\ldots a_r\}}
   C^{N,r}_{j_1\ldots j_{r-4}}(S)
\nn\\
&&+ \sum_{j_1,\ldots,j_{r-4}\in S}
   \left[g^{\cdot\cdot}g^{\cdot\cdot}g^{\cdot\cdot}%
   \Delta_{j_1\cdot}^{\cdot}\cdots\Delta_{j_{r-6}\cdot}^{\cdot}%
   \right]^{\{\mu_1\ldots\mu_r\}}_{\{a_1\ldots a_r\}}
   D^{N,r}_{j_1\ldots j_{r-6}}(S)
\nn\\
&&+\;\ldots   \;.\nn
\end{eqnarray}
The notation $[\cdots]^{\{\mu_1\cdots\mu_r\}}_{\{a_1\cdots a_r\}}$ 
stands for the distribution of the $r$ Lorentz indices $\mu_i$, and the momentum 
labels $a_i$,  to the vectors $\Delta_{j\,a_i}^{\mu_i}$ and metric tensors 
in all distinguishable ways. 
Note that the choice $r_N=0$, $a_i=N \,\,\forall \,i$ leads to the well known representation 
in terms of external momenta where the labels $a_i$ are not necessary, 
but we prefer a completely shift invariant notation here.

$S$ denotes an ordered 
set of propagator labels, corresponding to the momenta forming 
the kinematic matrix ${\cal S}$, defined by 
\bea
\calst_{ij} &=&  (r_i-r_j)^2-m_i^2-m_j^2\;, \;\quad i,j\in\{1,\ldots,N\}\;.
\label{eqDEFS}
\eea
The kinematic matrix ${\cal S}$ is related to the Gram matrix $G_{ij}$ 
($i,j=1,\ldots ,N-1$ for $r_N=0$)
by 
\be
\det G= (-1)^{N+1}B\,\det{\cal S}\;,\;B=\sum_{i,j=1}^N {\cal S}^{-1}_{ij}\;.\label{Bdef}
\ee
We should point out that the form factors of type $D^{N,r}_{j_1\ldots j_{r-6}}$ and beyond,
i.e. form factors associated with three or more metric tensors, 
are not needed for integrals where the rank $r$ does not exceed the number $N$ 
of propagators, no matter what the value of $N$ is. 
This is because integrals with $N\geq 6$ can be reduced algebraically to pentagons, 
without generating higher dimensional remainder terms, using the following formula recursively\,\cite{Binoth:2005ff}
\begin{displaymath}
I_N^{D,\mu_1\ldots\mu_r}(a_1,\ldots,a_r;S)  =  
 - \sum_{j \in S}{\cal C}_{ja_r}^{\mu_{r}} \, 
 I_{N-1}^{D,\mu_1\ldots\mu_{r-1}}(a_1,\ldots,a_{r-1};S\setminus \{j\})\quad (N\geq 6)\;,
%\label{TenRedNgeq6}
\end{displaymath}
where
${\cal C}_{j a}^{\mu}$ is the solution of the equation
\begin{equation}
\sum_{j \in S} {\cal S}_{ij}\, {\cal C}_{j\,a}^{\mu} = 
\Delta_{i\, a}^{\mu}
\;, \;\;\;\; a \in S
\;.
\label{eqCDEF}
\end{equation}
If $N \geq 7$, or in the case of 
exceptional kinematics, ${\cal S}$ is not invertible, so 
eq.~(\ref{eqCDEF}) does not have a unique solution. 
However, an explicit solution can be constructed 
as shown in\,\cite{Binoth:2005ff,Fleischer:2012hg}.
In this sense the tensor reduction of $N$-point integrals with $N\geq 6$ 
is trivial: integrals with $N\geq 6$ can be reduced iteratively to 
5-point integrals.
%, without generating higher dimensional remainders.
Therefore form factors for $N\geq 6$ are never needed.
If the rank $r$ of an $N$-point integral does not exceed $N$, 
Lorentz structures carrying at most two factors of $g^{\mu\nu}$
are sufficient, as the maximal form factor needed is $C^{5,5}_j$,
corresponding to a rank five 5-point integral.
However, for $r>N$, we can also have e.g. rank six 5-point integrals, 
which contain Lorentz structures involving three factors of $g^{\mu\nu}$,
and therefore form factors which go beyond the types $A^{N,r},B^{N,r},C^{N,r}$
are needed.

\medskip

The form factors are linear combinations of algebraic 
reduction coefficients, derived from the matrix ${\cal S}$, and $N$-point integrals
with $N\leq 4$. Explicit expressions for $r\leq N$ are given in~\cite{Binoth:2005ff}.

\subsection{Integrals}

The \GOLEMC{} program uses the fact that tensor integrals 
are related to  Feynman parameter integrals with Feynman parameters in the numerator.
A scalar integral, after Feynman parametrisation, can be written as 
\bea
I^D_N(S) &=& (-1)^N\Gamma(N-\frac{D}{2})\int \prod_{i=1}^N dz_i\,
\delta(1-\sum_{l=1}^N z_l)\,\left(R^2\right)^{\frac{D}{2}-N}\nn\\
&& R^2 =  
-\frac{1}{2} \sum\limits_{i,j=1}^N z_i\,\calst_{ij}  z_j\,\,-i\delta
\;.
\label{isca2}
\eea

The general relation between tensor integrals and parameter integrals 
with Feynman parameters in the numerator is
 well known~\cite{Davydychev:1991va,Bern:1992em,Binoth:1999sp}:
\bea
&&I^{D,\,\mu_1\ldots\mu_r}_N(a_1,\ldots,a_r\,;S) 
 =  
(-1)^r \sum_{m=0}^{[r/2]} \left( -\frac{1}{2} \right)^m\nn\\ 
&&
\sum_{j_1\cdots j_{r-2m}=1}^N \left[ 
 (g^{..})^{\otimes m}\,\Delta_{j_1\cdot}^{\cdot} \cdots \Delta_{j_r\cdot}^{\cdot}
\right]^{\{\mu_1\cdots\mu_r\}}_{\{a_1\cdots a_r\}}
\;
 I_N^{D+2m}(j_1 \ldots ,j_{r-2m}\,;S)\;,
\label{eq32}
\eea 
where $I_N^{D+2m}(j_1 \ldots ,j_{r-2m}\,;S)$ 
 is an integral with Feynman parameters in the numerator.
 $[r/2]$ stands for the nearest integer less or equal to $r/2$ and the symbol 
 $\otimes m$ indicates
 that $m$ powers of the metric tensor are present.
Feynman parameter integrals corresponding to 
diagrams where propagators $l_1,\dots,l_m$ are omitted 
%or {\it pinched},
with respect to the ``maximal" topology 
can be defined as
\bea
&&I^D_N(j_1,\dots,j_r;S\setminus \{l_1,\dots,l_m\}) =(-1)^N\Gamma(N-\frac{D}{2})
\nn\\
&& 
\int \prod_{i=1}^N dz_i\,
\delta(1-\sum_{k=1}^N z_k)\,
\delta(z_{l_1})\dots \delta(z_{l_m})z_{j_1}\dots z_{j_r}\left(R^2\right)^{D/2-N}\;.
\label{isca_pinch}
\eea

%\subsection{Basis integrals}\label{basisints}

The program \GOLEMC{} reduces the integrals internally to a set of 
basis integrals, i.e. the endpoints of the reduction 
(they do not form a basis in the mathematical sense, 
as some of them are not independent).
The choice of the basis integrals can have important effects on the numerical stability 
in certain kinematic regions.
Our reduction endpoints are 
4-point functions in 6 dimensions
$I_4^6$, which are IR and UV finite, 4-point functions in
$D+4$ dimensions, and various 2-point and 3-point functions, some of
the latter  with Feynman parameters in the numerator. This provides us with a 
convenient  separation of IR and UV divergences, as the IR poles are 
exclusively contained in
the triangle functions. 

Note that $I^{D+2}_3$ and $I^{D+4}_4$ are UV divergent, while  $I^{D}_3$
can be IR divergent. In the code, the integrals are represented as 
arrays containing the coefficients of their Laurent expansion in $\epsilon=(4-D)/2$.

If the endpoints of the reduction contain finite scalar four-point integrals,  
the latter have not all been coded explicitly, 
but some are automatically called from the library {\tt OneLOop}\,\cite{vanHameren:2010cp}.

%\subsection{Overview of the software structure}

%The structure of the \GOLEMC{} program is the following:
%There are four  main directories:
%\begin{enumerate}
%\item {\bf src:} the source files of the program
%\item {\bf demos:} some programs for demonstration
%\item {\bf tool:} scripts used to build the different parts of the program
%\item {\bf doc:} documentation which has been created with robodoc~\cite{robodoc} 
% robodoc --src ../src --doc . --multidoc --index --html
%\item {\bf test:} supplements the 
%demonstration programs, containing files to produce form factors with user-defined kinematics. 
%      The user can specify the rank, numerator, numerical point 
%      etc. via a steering file.
%\end{enumerate}
%The subdirectory structure is the same as described in \cite{Binoth:2008uq}.

%\section{Description of the individual software components}

%%%%%%%%%%%%%%%%%%%%%%%%%%%%%%%%%%%%%%%%%%%%%%%%%%%%%%%%%%%%%

\section{Extension of the program to higher rank}
\label{sec:highrank}

We focus here on the new features, for more details on the software 
components which are the same as in version 1.2, we refer to \cite{Binoth:2008uq,Cullen:2011kv}
and to the documentation contained in the program.

The call syntax for the higher rank form factors is analogous to the one
for $r\leq N$,
where the form factors defined in eq.\,(\ref{eq:formfactordef}) 
have been extended to include higher ranks. 
The code is built such that the reduction procedure is valid for arbitrary ranks.
Form factors up to rank ten for $N\leq 4$ and up to rank six 
for $N=5$ have been hardcoded, 
additional ones can be easily generated 
using the python script {\tt gen\_form\_factors.py} in the subdirectory {\tt tool/highrank}.
Explicit examples 
how to call the form factor for a rank four triangle, a rank five box 
and a rank six pentagon are given in section \ref{sec:examples}.

\begin{figure}
\begin{center}
\includegraphics[width=0.8\textwidth,page=2]{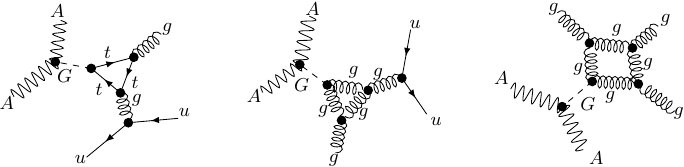}
\end{center}
\caption{Examples of diagrams with graviton exchange, where rank four 3-point and 
rank five 4-point integrals are needed. The graviton is denoted by $G$.}
\label{fig:graviton}
\end{figure}

\section{Installation}

The program can be downloaded as  golem95-1.3.0.tar.gz from the following URL:
{\tt http://golem.hepforge.org/95/}.
The installation instructions given below also can be found in the
{\tt README} file  coming with the code.
%Information and updates of the program can also be found at 
%{\tt http://projects.hepforge.org/golem/trac/wiki/golem95C}.

The installation setup is based on  autotools~\cite{Autotools}. 
To install the \GOLEMC{} library, type the following commands:\\
{\tt ./configure [--prefix=mypath] [--precision=quadruple] [FC=compiler]}\\
{\tt make}\\
{\tt make install}

The \texttt{--prefix} option denotes the installation prefix, under which
the directories \texttt{lib/} and \texttt{include/} are generated. If no
option is given, on a Linux system the configure script would choose
\texttt{prefix=/usr/local}. The argument \texttt{--precision} selects
double or quadruple precision to be used in the library; it should
be noted that quadruple precision is not supported by all Fortran compilers
and that \texttt{precision=double} is the default value.
If the variable \texttt{FC} is not set the first Fortran compiler which is
automatically detected will be used. Another variable commonly used is
\texttt{FCFLAGS} which allows one to pass compiler flags to the Fortran compiler.
%To run the demo file {\tt demos/demo\_LT.f},
%a  fortran77 compiler is also needed, which can be specified by the 
%install option {\tt F77=fortran77compiler}.

As an alternative to the call of finite box integrals from {\tt OneLOop}\,\cite{vanHameren:2010cp}, 
it is possible to call finite scalar box and triangle integrals 
with internal masses from {\tt LoopTools}\,\cite{Hahn:1998yk}. To use this option, the user 
should (a) install {\tt LoopTools}, and (b) use the option 
{\tt --with-looptools=path\_to\_libooptools.a} for the configure script, i.e. 
type the following commands:\\
{\small \tt ./configure [--prefix=mypath] [--with-looptools=path\_to\_libooptools.a]} 
{\tt [--precision=quadruple] [FC=compiler] [F77= fortran77compiler]}\\
{\tt make}\\
{\tt make install}.

\section{Usage and examples}
\label{sec:examples}

Examples for the usage of the program can be found in the subdirectory {\tt demos}. 
The examples for the higher rank form factors are also described below.
%in section~\ref{sec:examples}. 

The basic structure of a program using the calculation of 
$N$-point scalar integrals or form factors by \GOLEMC{} is:

\begin{flushleft}
{\tt call initgolem95(N)}\\
\vspace{2mm}
\dots fill kinematic matrix ${\cal S}$ \dots\\
\vspace{2mm}
{\tt call preparesmatrix()}\\
\vspace{2mm}
\dots evaluate integrals/form factors \dots\\
\vspace{2mm}
{\tt call exitgolem95()}
\end{flushleft}
\vspace{2mm}

A simple example for the calculation of a scalar three-point integral, which is evaluated by 
calling the form factor {\tt A30}, is given in Figure~\ref{fig:a30}.

{\tiny
\begin{lstlisting}[caption={Example of a program calling a scalar three-point integral.},label={fig:a30},language=fortran,showstringspaces=false,columns=flexible]
program main
  
  use precision_golem ! to get the type ki (for real and complex)
  use matrice_s       ! needed for initgolem95, s_mat, etc.
  use constante       ! contains useful constants 
  use form_factor_type
  use parametre       ! contains default parameter settings
  use form_factor_3p  ! module containing the three-point form factors 
  implicit none
  
  type(form_factor) :: res
  real(ki) :: s1, s2, s3, m1sq, m2sq, m3sq
  
  s1 = 1._ki
  s2 = 0._ki
  s3 = 0._ki
  m1sq =  2._ki
  m2sq =  0._ki
  m3sq =  0._ki

  call initgolem95(3) ! initialisation of caching system and 3x3 matrix S

  ! definition of the kinematic matrix S
  
  s_mat(1,:) = (/ -m1sq*2._ki , s2-m1sq-m2sq, s1-m1sq-m3sq /)
  s_mat(2,:) = (/ s2-m1sq-m2sq, -m2sq*2._ki , s3-m2sq-m3sq /)
  s_mat(3,:) = (/ s1-m1sq-m3sq, s3-m2sq-m3sq, -m3sq*2._ki  /)

  call preparesmatrix()
  
  ! call the scalar triangle
  
  res = a30(s_null)
    
  write (6,*) 'result='
  write (6,'("  1/epsilon^2 * (",e16.10,1x,"+ I*",1x,e16.10,")")') real(res%a,ki),aimag(res%a)
  write (6,'("+ 1/epsilon   * (",e16.10,1x,"+ I*",1x,e16.10,")")') real(res%b,ki),aimag(res%b)
  write (6,'("+ 1           * (",e16.10,1x,"+ I*",1x,e16.10,")")') real(res%c,ki),aimag(res%c)
  
  call exitgolem95()
  
end program main
\end{lstlisting}
}

\subsection{Three-point rank 4 example}
The executable for this example is created by {\tt make demo\_3point}.
The program \texttt{demos/}\hspace{0cm}\texttt{demo\_three\_point.f90} contains as an option 
the call to a rank four three-point form factor, $A34(1,1,2,3,S)$.
The integers in the argument list denote the labels of the corresponding 
momenta in the form factor representation, i.e. $A^{3,4}_{1123}$
in the notation of eq.~(\ref{eq:formfactordef}), denoting the coefficient of the term
$\Delta_{1\,a_1}^{\mu1}\Delta_{1\,a_2}^{\mu2}\Delta_{2\,a_3}^{\mu3}\Delta_{3\,a_4}^{\mu4}$ 
in the form factor decomposition of the 
Lorentz structure.
Note that using $a_j=3$ and $r_3=0$, $\Delta_{i\,a_j}$ is simply replaced by $r_i$.
The call syntax is the same as for the lower rank form factors.
The results are written to the file \texttt{test3point.txt}.
As a check for the user, the results to be obtained when calling $A34(1,1,2,3,S)$
are listed in {\tt table\_of\_results\_3point\_option\_n.txt}.\\
The option $n$ in  {\tt table\_of\_results\_3point\_option\_n.txt} denotes 
different choices for the kinematics of the three-point function.
Running the executable {\tt demo\_3point}, the user is prompted to give (a) the option for the kinematics
and (b) the option for the rank (or other features). 
The higher rank three-point example corresponds to option (b) number 7.

\subsection{Four-point rank 5 example}
The executable for this example is created by {\tt make demo\_4point}.
The program \texttt{demos/}\hspace{0cm}\texttt{demo\_four\_point.f90} contains as an option 
the call to a rank five four-point form factor, $A45(1,2,2,3,3,S)$.
The integers in the argument list denote the labels of the corresponding 
momenta in the form factor representation, i.e. $A^{4,5}_{12233}$
in the notation of eq.~(\ref{eq:formfactordef}), denoting the coefficient of the term
$\Delta_{1\,a_1}^{\mu1}\Delta_{2\,a_2}^{\mu2}\Delta_{2\,a_3}^{\mu3}\Delta_{3\,a_4}^{\mu4}\Delta_{3\,a_5}^{\mu5}$ 
in the form factor decomposition of the Lorentz structure.
% (with $r_4=0$).
The call syntax is the same as for the lower rank form factors.
The results are written to the file \texttt{test4point.txt}.
As a check for the user, the results to be obtained when calling $A45(1,2,2,3,3,S)$
are listed in 
{\tt table\_of\_results\_4point\_option\_n.txt}, where again $n$ denotes different 
choices for the kinematics. Once the kinematics is chosen, the rank five 4-point example 
corresponds to option (b) number 9.

\subsection{Five-point rank 6 example}
The executable for this example is created by {\tt make demo\_5point}.
The program \texttt{demos/}\hspace{0cm}\texttt{demo\_five\_point.f90} contains as an option 
the call to a rank four three-point form factor, $A56(1,1,2,3,4,5,S)$.
The integers in the argument list denote the labels of the corresponding 
momenta in the form factor representation, i.e. $A^{5,6}_{112345}$
in the notation of eq.~(\ref{eq:formfactordef}), denoting the coefficient of the term
$\Delta_{1\,a_1}^{\mu1}\Delta_{1\,a_2}^{\mu2}\Delta_{2\,a_3}^{\mu3}\Delta_{3\,a_4}^{\mu4}\Delta_{4\,a_5}^{\mu5}\Delta_{5\,a_6}^{\mu6}$ 
in the form factor decomposition of the Lorentz structure.
The call syntax is the same as for the lower rank form factors.
The results are written to the file \texttt{test5point.txt}.
As a check for the user, the results to be obtained when calling $A56(1,1,2,3,4,5,S)$
(option 5 in \texttt{demo\_five\_point.f90}) are listed in 
{\tt table\_of\_results\_5point.txt}.

\section{Conclusions}
We have presented an extension of the program  \GOLEMC \  which 
provides a  library of scalar and tensor integrals 
where the tensor rank $r$, denoting the number of loop momenta in the numerator, 
can exceed the number of propagators $N$.
We have implemented ranks up to $r=10$ for  $N \leq 4$, and
5-point and 6-point functions up to rank $r= N+1$.
The use of both real or complex masses is possible within the same setup.
The program, which is an extension of on an earlier version of the 
{\tt golem95C} library, now also can be used in the presence of 
effective vertices and in models  where the presence of spin two particles  
can require integrals with higher ranks than usually needed in
renormalizable theories.
Due to an appropriate interface, the program can  be used both within  
a traditional tensor reduction approach as well as within 
unitarity-inspired numerical reconstruction of the integrand at the tensorial level.
The program is publicly available at 
{\tt http://golem.hepforge.org/95/}. 

\section*{Acknowledgements}
We would like to thank the {\sc GoSam} collaboration 
for feedback  on \GOLEMC{}, and in particular 
Gavin Cullen for numerous tests and helpful discussions,
Edoardo Mirabella  for useful comments, and Hans van Deurzen, Gionata Luisoni 
and Sadok Zidi for comparisons.
%We also acknowledge the support of the Research Executive Agency (REA)
%of the European Union under the Grant Agreement number
%PITN-GA-2010-264564 (LHCPhenoNet).

%\section{Appendix}
\renewcommand \thesection{\Alph{section}}
\renewcommand{\theequation}{\Alph{section}.\arabic{equation}}
\setcounter{section}{0}
\setcounter{equation}{0}
\section{Appendix: $\tilde{k}$ - integrals }
\label{sec:appendix}

Here we give a list of integrals which are needed if the momenta in the numerator 
are split into a 4-dimensional and a ($D-4$)-dimensional part, according to 
$k_{(D)}^{\mu}  =  \hat{k}_{(4)}^{\mu} + \tilde{k}_{(-2\eps)}^{\mu}, 
k_{(D)}^2  =  \hat{k}^2 + \tilde{k}^2$. 
Using  $\hat{q}_a=\hat{k}+r_a$, where $\hat{k}$ is the loop momentum in 4 dimensions, 
and $r_a$ is a combination of external momenta, we define 
\begin{equation}
\label{eq:ktilde}
I_N^{D,\alpha;\mu_1\ldots\mu_r}(a_1,\ldots a_r; S)\equiv%nl
\int\!\!\frac{\diff[D]k}{i\pi^{D/2}}\frac{\left(\tilde{k}^2\right)^\alpha
\hat{q}_{a_1}^{\mu_1}\cdots \hat{q}_{a_r}^{\mu_r}}{%nl
\prod_{j=1}^N(q_j^2-m_j^2+i\delta)}\text{.}
\end{equation}
The integrals with additional powers of $(\tilde{k}^2)^\alpha$ in the numerator
are related to integrals in higher dimensions by 
\begin{multline}
\label{eq:TItoSI}
I_N^{D,\alpha;\mu_1\ldots\mu_r}(a_1,\ldots,a_r; S)=
(-1)^{r+\alpha}\frac{\Gamma(\alpha+D/2-2)}{\Gamma(D/2-2)}\sum_{m=0}^{\lfloor r/2\rfloor}%nl
\left(-\frac12\right)^m\times\\\sum_{j_1,\ldots,j_{r-2m}=1}^N [\underbrace{%nl
\hat{g}^{\bul\bul}\cdots \hat{g}^{\bul\bul}}_m%nl
\Delta_{j_1\bul}^{\bul}\cdots\Delta_{j_{r-2m}\bul}^{\bul}%nl
]^{\mu_1\ldots\mu_r}_{a_1\ldots a_r}%nl
I_N^{D+2\alpha+2m}({j_1},\ldots,{j_{r-2m}};S)\text{,}
\end{multline}
where $\Delta_{ij}^{\mu}=r_i^{\mu}-r_j^{\mu}$.
Note that for $D=4-2\eps$, the prefactor in eq.~(\ref{eq:TItoSI}) reads 
\be
\frac{\Gamma(\alpha+D/2-2)}{\Gamma(D/2-2)}=
\frac{\Gamma(\alpha-\eps)}{\Gamma(-\eps)}=-\eps\,(\alpha-1)!+{\cal O}(\eps^2)\;.
\ee
Therefore, integrals with $\alpha>0$ will only contribute if the integrals 
$I_N^{D+2\alpha+2m}$ are UV divergent, 
since we can drop terms of ${\cal O}(\eps)$ for one-loop applications.
The  coefficient of the UV pole of these integrals, which is projected out in this way, 
will contribute to the so-called ``rational part" of an amplitude.
For $D=4-2\eps$, the integral $I_N^{D+2\alpha+2m}$ will 
be proportional to $\Gamma(\eps-\eta)$, with $\eta=2-N+\alpha+m$.
Therefore it will contain an UV divergence if $\eta\geq 0$.

%\bea
%\int \frac{d^{D}k}{i \pi^{\frac{D}{2}}}
% \,(\tilde{k}^2)^\alpha \,f(k^\mu,k^2)  & = &(-1)^\alpha
% \frac{\Gamma(\alpha+\frac{D}{2}-2)}{\Gamma(\frac{D}{2}-2)}
% \int \frac{d^{D+2\alpha}k}{i \pi^{\frac{D}{2}+\alpha}} f(k^\mu,k^2)\;.
% \label{ktilde}
%\eea
The results for those integrals which are relevant for the rational part
can be given in a general form~\cite{Reiter:2009kb}.
\begin{eqnarray}
&&\eps I_N^{D+2\alpha+2m}(l_1,\ldots,l_r;S)=
\frac{(-1)^N}{2^\eta\eta!}\sum_{j_1,\ldots j_{2\eta}=1}^N {\cal S}_{j_1j_2}\cdots {\cal S}_{j_{2\eta-1},j_{2\eta}}
P_N(l_1,\ldots,l_r,j_1,\ldots,j_{2\eta})\nn\\
&&\eta=2-N+\alpha+m\text{,}
\label{ktilgeneral}
%&&\eps I_N^{D-4+2N}(l_1,\ldots,l_r;S)=(-1)^NP_N(l_1,\ldots,l_r)\text{,}\\
%&&\eps I_N^{D-4+2(N+1)}(l_1,\ldots,l_r;S)=
%\frac{(-1)^N}{2}\sum_{j_1,j_2=1}^N S_{j_1j_2}\,P_N(j_1,j_2,l_1,\ldots,l_r)\text{,}
\end{eqnarray}
where we define 
\begin{equation}\label{eq:solvektil}
P_{t_1,t_2,\ldots,t_N}=
\frac{\prod_{j=1}^N(t_j!)}{(N-1+\sum_{i=1}^N t_i)!}
\end{equation}
to arrive at $P_N(j_1,\ldots,j_s)$,
which counts the indices in an expression:
\begin{equation}
P_N(j_1,\ldots,j_s)=P_{\left(\sum_{i=1}^s\delta_{1,j_i}\right),\ldots,\left(\sum_{i=1}^s\delta_{N,j_i}\right)}\text{.}
\end{equation}
Further, ${\cal S}_{j_1j_2}$  denotes an element of the kinematic matrix ${\cal S}$.

The integrals $I_N^{D+2\alpha+2m}$ will be UV divergent for $2\alpha+2m\geq 2N-4$.
Below give the relevant explicit expressions, derived from eq.~(\ref{ktilgeneral}).
Terms which will be of order ${\mathcal O}(\eps)$ are dropped.
\begin{subequations}
\begin{align}
&\eps I_N^{D-4+2N}(S)=\frac{(-1)^N}{(N-1)!}\displaybreak[1]\\
&\eps I_N^{D-4+2N}(l_1;S)=\frac{(-1)^N}{N!}\displaybreak[1]\\
&\eps I_N^{D-4+2N}(l_1, l_2;S)=
\frac{(-1)^N}{(N+1)!}\left(1+\delta_{l_1l_2}\right)\displaybreak[1]\\
&\eps I_N^{D-4+2N}(l_1, l_2,l_3;S)=\frac{(-1)^N}{(N+2)!}\\
&\qquad\times\left(1+\delta_{l_1l_2}+\delta_{l_1l_3}
+\delta_{l_2l_3}+2\delta_{l_1l_2}\delta_{l_2l_3}\right)\nonumber\displaybreak[1]\\
&\eps I_N^{D-4+2N}(l_1, l_2,l_3,l_4;S)=\frac{(-1)^N}{(N+3)!}\\
&\qquad\times(\delta_{l_1l_2}(6\delta_{l_1l_3}\delta_{l_2l_4}+2\delta_{l_1l_3}
+2\delta_{l_2l_4}+\delta_{l_3l_4})\nonumber\\
&\qquad\quad+2\delta_{l_3l_4}(\delta_{l_1l_3}+\delta_{l_2l_4})
+\delta_{l_1l_3}\delta_{l_2l_4}+\delta_{l_1l_4}\delta_{l_2l_3}\nonumber\\
&\qquad\quad+\delta_{l_1l_2}+\delta_{l_1l_3}+\delta_{l_1l_4}
+\delta_{l_2l_3}+\delta_{l_2l_4}+\delta_{l_3l_4}+1)\nonumber\displaybreak[1]\\
&\eps I_N^{D-4+2(N+1)}(S)=\frac{(-1)^N}{2(N+1)!}
\left(\sum_{j_1,j_2=1}^N{\cal S}_{j_1j_2}+\tr{{\cal S}}\right)\displaybreak[1]\label{r6box}\\
&\eps I_N^{D-4+2(N+1)}(l_1;S)=\frac{(-1)^N}{2(N+2)!}\sum_{j_1,j_2=1}^N{\cal S}_{j_1j_2}\left(1+\delta_{j_1j_2}\right)\\
&\qquad\times\left(1+\delta_{l_1j_1}
+\delta_{l_1j_2}\right)\nonumber\displaybreak[1]\\
&\eps I_N^{D-4+2(N+1)}(l_1,l_2;S)=
\frac{(-1)^N}{2(N+3)!}\sum_{j_1,j_2=1}^N{\cal S}_{j_1j_2}\\
&\qquad\times(\delta_{j_1j_2}(6\delta_{j_1l_1}\delta_{j_2l_2}
+2\delta_{j_1l_1}+2\delta_{j_2l_2}+\delta_{l_1l_2})\nonumber\\
&\qquad\quad+2\delta_{l_1l_2}(\delta_{j_1l_1}+\delta_{j_2l_2})
+\delta_{j_1l_1}\delta_{j_2l_2}+\delta_{j_1l_2}\delta_{j_2l_1}\nonumber\\
&\qquad\quad+\delta_{j_1j_2}+\delta_{j_1l_1}+\delta_{j_1l_2}
+\delta_{j_2l_1}+\delta_{j_2l_2}+\delta_{l_1l_2}+1)\;.\nonumber
\end{align}
\end{subequations}
We also give here the explicit formulas for rank 6 pentagon integrals involving $\tilde{k}^2$ terms, 
because the latter are special to higher rank extensions.
Again, terms of ${\cal O}(\eps)$ are dropped.
\bea
&&I_5^{D,3}(S)=
\int\!\!\frac{\diff[D]k}{i\pi^{D/2}}\frac{\left(\tilde{k}^2\right)^3
}{%nl
\prod_{j=1}^5(q_j^2-m_j^2+i\delta)}=-\frac{1}{12}\\
&&I_5^{D,2;\mu_1 \mu_2}(a_1,a_2; S)=
\int\!\!\frac{\diff[D]k}{i\pi^{D/2}}\frac{\left(\tilde{k}^2\right)^2\;
\hat{q}_{a_1}^{\mu_1}  \hat{q}_{a_2}^{\mu_2}}{%nl
\prod_{j=1}^5(q_j^2-m_j^2+i\delta)}=-\frac{1}{48}\,g^{\mu_1\mu_2}\\
&&I_5^{D,1;\mu_1\cdots \mu_4}(a_1,\ldots,a_4; S)=
\int\!\!\frac{\diff[D]k}{i\pi^{D/2}}\frac{\tilde{k}^2\;
\hat{q}_{a_1}^{\mu_1} \ldots \hat{q}_{a_4}^{\mu_4}}{%nl
\prod_{j=1}^5(q_j^2-m_j^2+i\delta)}\nn\\
&&\qquad =-\frac{1}{96}\,
\left[g^{\mu_1\mu_2}g^{\mu_3\mu_4}+ g^{\mu_1\mu_3}g^{\mu_2\mu_4}
+g^{\mu_1\mu_4}g^{\mu_2\mu_3}\right]\;.
\eea

As another example, we give the expressions for the rational part of box integrals in $D+6$ dimensions, 
needed for rank six four-point functions,
which can be obtained from ~\eqref{r6box}:
\be
\eps I_4^{D+6}(S)=\frac{1}{240}\left(\sum_{i,j=1}^4 (\Delta^2_{ij}-m_i^2-m_j^2)-2\sum_{i=1}^4 m_i^2\right)
+{\cal O}(\eps)\;.
\ee

%\begin{thebibliography}{99}
%\bibliography{golem}

%\end{thebibliography}

\end{document}